\documentclass[twocolumn,amsmath,nofootinbib,aps]{revtex4}

\usepackage{amssymb}
\usepackage{amsmath}
\usepackage{epsfig}
\usepackage{subfigure}
\usepackage{mathrsfs}
\usepackage{longtable}

\newcommand{\be}{\begin{equation}}
\newcommand{\ee}{\end{equation}}
\newcommand{\bea}{\begin{align}}
\newcommand{\eea}{\end{align}}

\newcommand{\unit}[1]{\ensuremath{\, \mathrm{#1}}}

\begin{document}

\title{The Axiverse Extended: Vacuum Destabilisation, Early Dark Energy and Cosmological Collapse}
\author{David J. E. Marsh}
\affiliation{Rudolf Peierls Centre for Theoretical Physics, University of Oxford, 1 Keble Road, 
Oxford, OX1 3NP, UK}


\begin{abstract}
A model is presented in the philosophy of the ``String Axiverse'' of \cite{axiverse2009} that incorporates a coupling of ultralight axions to their corresponding moduli through the mass term. The light fields roll in their potentials at late times and contribute to the dark sector energy densities in the cosmological expansion. The addition of a coupling and extra field greatly enrich the possible phenomenology of the axiverse. There are a number of interesting phases where the axion and modulus components behave as Dark Matter or Dark Energy and can have considerable and distinct effects on the expansion history of the universe by modifying the equation of state in the past or causing possible future collapse of the universe. In future such a coupling may help to alleviate fine tuning problems for cosmological axions. We motivate and present the model, and briefly explore its cosmological consequences numerically.
\end{abstract}

\maketitle
\section{Introduction}
\label{intro}

\subsection{The Mass Scale Of String Axions}

It was proposed in \cite{axiverse2009} that a generic prediction of string theory is the existence of many light axions. The Lagrangian of such an axion can be characterised using two parameters: the symmetry breaking scale, $f_a$ (also referred to as the axion decay constant), and the overall scale of the potential, $\Lambda$, appearing in the effective four dimensional Lagrangian:
\begin{equation}
\mathcal{L} = \frac{f_a^2}{2}(\partial\theta)^2 - \Lambda^4 U(\theta)
\end{equation}
where $U(\theta)$ is some periodic potential. Bringing the kinetic term into canonical form we define the field $\phi = f_a \theta$, with Lagrangian:
\begin{equation}
\mathcal{L} = \frac{1}{2}(\partial \phi)^2 - V_{ax}(\phi)
\label{eqn:lagrangian}
\end{equation}
where $V_{ax}(\phi)$ is again a periodic potential. Expanding the potential to quadratic order we find that the mass is given by:
\begin{equation}
m_a^2 = \frac{\Lambda^4}{f_a^2}
\end{equation}
Hence, any axion is equivalently parameterised by its mass and symmetry breaking scale. The authors of \cite{axiverse2009} then go on to assert that, because of the scaling of $m_a$ and $f_a$ with the parameters in a generic string theory compactification, and the complexity of such compactifications, that one should expect $f_a$ to remain roughly fixed at some high scale $f_a\sim 10^{16}\unit{GeV}$, whilst $m_a$ should distribute roughly evenly on a logarithmic mass scale all the way down to the Hubble scale today of $H_0\sim 10^{-33}\unit{eV}$.

The argument can be summarised as follows. Axions arise from the existence of closed cycles in the compact space: one axion for each. The symmetry breaking scale, $f_a$, scales inversely with the action, $S$, due to non-perturbative physics on the corresponding cycle:
\begin{equation}
f_a \sim \frac{M_{pl}}{S}
\end{equation}
The action then typically scales with the \emph{area} of the corresponding cycle, so that across all axions in a given compactification volume $f_a$ will not vary over many orders of magnitude. The standard arguments then lead to stringy values of $f_a\sim 10^{16}\unit{GeV}$~\cite{witten2006}. The crucial observation that enables the assertion about the distribution of axion masses to be made is the exponential dependence of the scale $\Lambda$ on $S$:
\begin{equation}
\Lambda^4 = \mu^4 e^{-S}
\end{equation}
where $\mu$ is a mass scale: in the axiverse it is the geometric mean of the supersymmetry (SUSY) breaking scale and the string/Planck scale. Therefore, as the areas of cycles scale over the volume of the compact space, and given that even the simplest compactifications contain hundreds of closed cycles of different orders, we can expect axion masses to distribute roughly evenly on a log scale.

It should finally be noted that in the axiverse scenario the lightest axions are \emph{not} the standard QCD axion \cite{pecceiquinn1977}, however they do owe their existence to it. String axions can be removed from the spectrum of light fields by tree level liftings towards the string/Planck scale, however if it is to solve the strong CP problem the QCD axion must escape such liftings. It would then be considered anomalous if the QCD axion were the only axion to escape such a lifting and remain light.

\subsection{String Axions As Dark Matter}

Axions and other ultralight scalar fields are a well motivated candidate to make up a proportion of the dark matter in our universe \cite{hu2000,amendola2005,marsh2010}, and have also been motivated in many works to solve the problem of dark energy (see, for example, \cite{panda2010,catena2007}). Studies of the QCD axion \cite{pecceiquinn1977} (see \cite{dine2000} for a review of the strong CP problem and the axion as a possible solution) as a dark matter candidate \cite{ipser1983,turner1983,abbott1983,preskill1983,turner1986,berezhiani1992,rosenberg1999,fox2004} often depended heavily on the thermal properties of the axion, particularly its temperature dependent mass. However, the thermal properties are specific to the non-perturbative instanton physics of QCD. In contrast, when we study axions arising in a generic context from string theory \cite{witten2006,axiverse2009} the non-perturbative physics of the axion potential could arise from a variety of sources, and as such it is common to work with a simplified form of the potential with no temperature dependence. Therefore, cosmological axions in such a scenario have relatively simple dynamics: they acquire some initial conditions through spontaneous symmetry breaking of the underlying global Peccei-Quinn (PQ) symmetry at the scale $f_a$, and the field remains frozen at this value until the mass (provided by the potential due to non-perturbative physics, which arises at the scale $\mu$. In the QCD case this happens near $\Lambda_{QCD}$) overcomes the Hubble friction, at which point the axion begins rolling towards and oscillating about its minimum in the potential. Such fields, which acquire only small masses due to explicit breaking of a symmetry due to non-perturbative effects (instantons breaking the remaining shift symmetry of the PQ axion in QCD) are known as pseudo-Nambu-Goldstone bosons~\cite{hill1988}, and are the common motivation for the study of ultralight fields in cosmology.

As a cosmological fluid the axions are purely gravitationally coupled: whilst frozen they behave as a cosmological constant, and then make a short transition to oscillatory behaviour before behaving as a pressureless cold dark matter (CDM) component. The fraction of the total energy density in axions resulting from this non-thermal production depends upon the initial value the field acquires after symmetry breaking and its distance from the potential minimum (the misalignment angle), with a negligible thermal component from the (Planck and symmetry breaking scale-) suppressed couplings of axions to the standard model via higher dimensional operators. In such a scenario the initial misalignment angle therefore often requires tuning to produce a cosmology consistent with observations if we assume the existence of an axion with a certain mass. The tuning may be to small values of the misalignment angle if we wish not to overclose the universe with heavier axions, or to large values of the misalignment angle if we would like the dark matter to contain some significant fraction of ultralight axions \cite{marsh2010}. It should be noted, however, that the measure of fine tuning in such scenarios depends not only on the potential, which is unknown and can in some cases be non-periodic~\cite{panda2010} (in which case the concept of tuning is more ill-defined), but also on the details of inflationary physics~\cite{linde1991,tegmark2006,hertzberg2008,mack2009a,mack2009b}. 

These fine tunings are a problem and a blessing for ultralight fields. In \cite{marsh2010} the ``anthropic boundary'' for ultralight axions was discussed. This refers to the region of axion parameter space where it is no longer possible for the axions to make up an $\mathcal{O}(1)$ fraction of the dark matter, due to the periodic nature of the field making $\phi=f_a\pi$ the maximum possible misalignment. Unless the periodicity is broken as was done in \cite{panda2010} and assumed in \cite{marsh2010}, or anharmonic effects in the potential are accounted for \cite{turner1986}, ultralight axions cannot make up large enough fractions of the dark matter to be readily observable, but consequently pose no danger of ``overclosing'' the universe and producing a Hubble rate inconsistent with observations. Part of the motivation for this work is to provide a mechanism that decouples the axion density somewhat from its initial misalignment angle via a tracking mechanism \cite{ferreira1998,copeland1998}, which may allow for observable densities in light axions, or to accommodate larger $f_a$ for the heavier axions.

Recently a specific construction in string theory realising the ``axiverse'' of \cite{axiverse2009} was given~\cite{acharya2010a,acharya2010b}. The authors of \cite{acharya2010a,acharya2010b} suggested that non-thermal processes are particularly important for the cosmology of such a model. In this paper I will propose a simple extension of the axiverse that includes further non-thermal processes that couple the axion fluid and greatly enrich the phenomenology of axion dark matter in a way analogous to the enrichment of dark energy phenomenology arrived at through the study of coupled quintessence~\cite{amendola2000}. In this framework the behaviour of the axion dark matter component is no longer a simple transition, and one hopes to look for novel features in the late time expansion history of the universe, as has been done fruitfully in many works on dark energy (see, for example, \cite{linder2010,calabrese2010}).

The perturbations of ultralight dark matter axions and their effect on structure formation has been studied in many previous works~\cite{marsh2010,amendola2005}, the basic result being that the large Compton wavelength gives rise to a quantum pressure and suppresses structure formation below that scale~\cite{hu2000}, in a way largely analogous to the presence of a massive neutrino~\cite{amendola2005}. The study of perturbations in this extended scenario will be left to a future work.

\subsection{Moduli}

Moduli are scalar fields in string theory that control the size and shape of the compact manifold. That there are many of these and they can contribute significantly to the cosmological energy density if they are allowed to roll in their potentials is known as the ``cosmological moduli problem'', and moduli stabilisation is an important problem in string theory, related to the landscape, the existence of stable vacua, and the cosmological constant problem (see \cite{conlon2006} and references therein). A typical potential for moduli stabilisation in the LARGE volume scenario for a modulus $\chi$ is of the parametric form:
\begin{equation}
V_{mod}(\chi) =  B e^{-2C\chi} - D e^{-C\chi}
\label{eqn:modpot}
\end{equation} 
where $B$, $D$ and $C$ are for our purposes free parameters, with the only restriction that for a stabilising minimum we have $2B>D$. The typical scales of these parameters will be discussed shortly. We will also add an arbitrary (cosmological) constant, $\Lambda$ (not to be confused with the scale of the axion potential), to the potential $V_{mod}$ to account for the frozen/vacuum expectation values of other moduli and standard model fields, the vacuum energy itself, and the bare gravitational $\Lambda$ in Einstein's equations, thus allowing for a late time de-Sitter expansion and acceptable phenomenology. This is a prescription necessary in most all moduli stabilisation, since the vacuum found is always anti-de-sitter. 

The modulus dynamics are therefore expected to be a small perturbation about the standard model for dark energy, similar to models of varying equation of state, quintessence or Early Dark Energy (EDE) (For a comprehensive review of dark energy theory and phenomenology, see \cite{copeland2006}.). This prescription, like many cosmological models, is  not answering the cosmological constant problem, only giving it some alternative phenomenology. The fate of the universe given this scenario is discussed in Section~\ref{results}.

In the LARGE volume moduli stabilisation scenario of~\cite{conlon2006,conlon2007} the lightest modulus is the volume modulus, which like the dilaton controls the overall scale of couplings in the model, but  we can expect the presence of \emph{many light moduli}. There are moduli corresponding to each closed cycle in the compact space, and as such there is a modulus for each axion. In particular there will be a modulus controlling the area of a cycle and as such:
\begin{equation}
S \sim \chi
\end{equation}
The modulus mass scale that leads to rolling in the potential is often tied to the axion mass, so that the presence of cosmologically rolling axions suggests the presence of \emph{cosmologically rolling moduli}. In this way we extend the axiverse.

The displacement of moduli from their stabilised values (vacuum destabilisation) by astrophysical processes is not a new idea: it has been exploited before in chameleon models (where destabilisation is attained via coupling to density) and other similar scenarios (see, for example \cite{conlon2010,khoury2004,brax2004,brax2010}). The variant on the scenario that is proposed here is as follows. The lightest axions and moduli have masses below the Hubble scale and are stabilised by Hubble friction, contributing to an effective cosmological constant. The heaviest moduli are stabilised by potentials of the form Eqn.~\ref{eqn:modpot}, which includes the moduli for standard model couplings \footnote{We will not investigate any evolution of the couplings with cosmic scale. The possibility and implication of such effects in this model will be the subject of future work}. There are then cosmological axions that roll in their potentials at late time and contribute to the dark matter density and perturbations in a distinct way from standard CDM. As such, \emph{some} moduli \emph{may} also roll in their potentials on cosmological time scales, which will lead to a rolling of the scale of the axion potential and consequently a \emph{rolling of the axion mass}. It is this feature of coupling that we hope to exploit in looking for novel features in the cosmological expansion rate and equation of state in the dark sector. It is worth stressing again that cosmologically rolling moduli are not general, but we explore the phenomenology of allowing such a scenario for some of the moduli. 

A final word of warning should be made about the addition of cosmologically relevant scalars to any model. Without a symmetry, such as the axion shift symmetry, which forbids scalar couplings directly to such terms as $F_{\mu\nu}F^{\mu\nu}$ (although they might be induced at loop order like the axion-photon coupling), scalars will always appear multiplying such terms and if they are light will induce long range, gravitational strength ``fifth forces''. This problem is generic to scalar quintessence models, though often ignored. The chameleon models are geared towards solving this problem \cite{brax2007}. This work will not address such fifth force constraints, but they should be born in mind when making any detailed analysis, and will also be the subject of future investigation.

\subsection{Comments}

In this model we are working in the Leibnizian/Panglossian philosophy of the authors of \cite{axiverse2009} to use the vastness of  the string landscape to look for general properties exploitable for cosmological phenomenology i.e. realistic string compactifications and scenarios for the moduli are complicated and varied but have many general and model independent features \cite{conlon2007}. In the literature there has been a long standing link between string theory and inflation physics (see~\cite{balasubramanian2007} and references therein). In particular it was the aim of \cite{balasubramanian2007} to connect inflationary observables to topological properties of the compact space; this is analogous to ``cohomologies from cosmology'' in \cite{axiverse2009}. However in the simplest axiverse scenario for dark matter axions there are two parameters per axion that can be constrained using cosmology: the axion mass, $m_a$, and the axion fraction, $f = \Omega_a/\Omega_m$. The fraction $f$ is determined by the initial misalignment angle and tells us nothing about the compact space or vacuum, so any cosmological bounds based on the expansion rate or formation of structure can only limit the contribution from a given axion or number of axions  of given masses rather than place definite constraints on the number and size of closed cycles that determine the masses. There could be many light axions of many masses but if their contribution to the energy density is too small then we will not observe them \footnote{The existence of an axion field coupling to standard model fields can be constrained in many other ways: for example, by light shining through a wall experiments (for a recent review, see~\cite{redondo2010}), and astrophysical processes such as those explored in \cite{mortsell2003,ostman2005,arvanitaki2010,rosa2010}}. An observation of the effects of a number of axions, modulo a prior on the potential and degree of fine tuning, may also hope to place bounds on their common axion decay constant, $f_a$. Assumptions about inflation can also bound $f_a$, as discussed in \cite{linde1991,tegmark2006,hertzberg2008,mack2009a,mack2009b}, and vice versa an assumed ultralight axion bounds $H_I$ from considerations of isocurvature perturbations, as discussed in \cite{fox2004}.

Considering axions and moduli in the landscape and whether they survive to be observed is closely linked to the question of SUSY breaking, and is discussed in~\cite{dine2010}. We will discuss it no further here, except to stress again that issues of naturalness and fine tuning \emph{within} the landscape will not concern us here: we simply motivate a theoretically plausible, phenomenolgically viable, and observationally testable scenario.

Extending the axiverse scenario as proposed here gives much greater scope for direct connections between late time cosmology and beyond the Standard Model/string physics, akin to those already fruitfully explored in inflationary physics, but distinct from those connections already explored in tackling the dark matter problem with weakly interacting massive particles (WIMPs) in the context of supersymmetric extensions of the standard model (the MSSM and its progeny), or in directly addressing the cosmological constant problem. However, as phenomenologists we must be careful about any statements we make about fundamental physics based on any results obtained using our models. In a parameterised phenomenological model such as the one proposed here, the parameters should be taken as just that: it is only in the context of a full string model such as those in~\cite{witten2006,conlon2006,balasubramanian2007,acharya2010a} (with all the assumptions that go into constructing such a model) that the parameters take on their physical high energy physics meanings.

\section{The Model: Coupled Axions in the Dark Sector}
\label{model}

\subsection{Equations of Motion}

The coupled axion-modulus Lagrangian takes the following form:
\begin{align}
\mathcal{L} &= \frac{1}{2}(\partial \phi)^2 + \frac{1}{2}(\partial \chi)^2 - V(\phi,\chi)  \nonumber \\
V(\phi,\chi) &= B e^{-2C\chi} - D e^{-C\chi} + \frac{1}{2}e^{- \tilde{C} \chi}M^2 \phi^2  \nonumber \\ 
\label{eqn:lagrangiantotal}
\end{align}
where $\phi$ is the axion, $\chi$ is the modulus, $B$, $D$ are dimension four parameters for the modulus potential, $C$ is related to the overall volume of the compact space in string units, $\tilde{C}$ is related to the instanton action, $M^2 = \mu^4/f_a^2$ and we make the simplifying assumption to work with only the mass term for the axion. In the interests of economy of parameters we will often take $\tilde{C} = C$. Also, when $C \neq \tilde{C}$ the problem of minimising the potential becomes much less tractable generically, and for specific values of $C$, $\tilde{C}$ many minima appear (e.g. 19 minima with $C=10$, $\tilde{C}=1$), which is related to the emergence of the landscape in string theory and the analysis of which is beyond the scope of this work.

In this work we will only be concerned with the homogeneous background fields, so will use the notation $\phi \equiv \phi_0 = \phi_0(\tau)$, where $\tau$ is conformal time, and similarly for the modulus. 

The energy momentum tensor and equations of motion for the coupled system follow in the usual way from the Lagrangian. For a homogeneous Friedmann-Robertson-Walker metric, with scale factor $a$, in conformal time:
\begin{align}
\ddot{\phi} + 2\mathcal{H} \dot{\phi} + e^{-C\chi}\frac{\mu^4}{f_a^2} a^2\phi &= 0  \nonumber \\
\ddot{\chi} + 2 \mathcal{H}\dot{\chi} - C a^2 (2B e^{-2C\chi} - D e^{-C \chi})  &= \frac{1}{2}C \frac{\mu^4}{f_a^2} a^2 e^{-C\chi} \phi^2 \nonumber \\
\end{align}
where over dots denote derivatives with respect to conformal time, and $\mathcal{H}=\dot{a}/a$. The energy momentum tensor for the combined axion-modulus system has the form of a perfect fluid with energy density $\rho$ and pressure $P$: $T^0_{\ \ 0} = -\rho$, $T^i_{\ \ j} = P\delta^i_{\ \ j}$. This gives:
\begin{align}
\rho &= \frac{a^{-2}}{2}(\dot{\phi}^2 +\dot{\chi}^2) + V(\phi,\chi)  \nonumber \\
P &= \frac{a^{-2}}{2}(\dot{\phi}^2 +\dot{\chi}^2) - V(\phi,\chi) \nonumber \\
\label{eqn:density}
\end{align}
Due to the coupling, only these combined quantities obey the conservation equation $\dot{\rho} = -3 \mathcal{H}(\rho + P)$. 

The form of the potential suggests a natural splitting of this into components due to the axion, subscript $\phi$, and modulus, subscript $\chi$:
\begin{align}
\rho_\phi &= \frac{a^{-2}}{2}\dot{\phi}^2 + \frac{1}{2}e^{-\tilde{C} \chi}M^2\phi^2  \nonumber \\
P_\phi &= \frac{a^{-2}}{2}\dot{\phi}^2 - \frac{1}{2}e^{-\tilde{C} \chi}M^2\phi^2  \nonumber \\
\rho_\chi &= \frac{a^-2}{2}\dot{\chi}^2 + Be^{-2C\chi}-De^{-C\chi} \nonumber \\
P_\chi &= \frac{a^-2}{2}\dot{\chi}^2 - Be^{-2C\chi}+De^{-C\chi} \nonumber \\
\label{eqn:rhosplit}
\end{align}
though in certain cases this distinction should be looked at more carefully~\cite{beyer2010}.

With these definitions we will investigate the scalings of the energy density for axion and modulus components, and also the combined system. It is not only the scaling of the energy density that effects the cosmological expansion history: we will also investigate the equation of state, given by: $w_i = P_i/\rho_i$ for components $i = ax,mod,ax+mod,tot$, which will effect the expansion rate in the usual way~\cite{book:dodelson,book:peacock}.

The Friedmann equation is:
\begin{equation}
\mathcal{H}^2 = \frac{8\pi G}{3}a^2\rho
\end{equation}
In addition to the axion-modulus system of Eqn.~\ref{eqn:density} we will consider components of the energy density coming from radiation, $\rho_\gamma$, CDM, $\rho_c$, and a cosmological constant, $\rho_\Lambda$, all of which will redshift in the usual way. This now completes the description of the system.

\subsection{The Scales of Parameters}

This model is to be understood phenomenologically, and thus all the parameters will be taken as free when searching for interesting cosmological features, however it will be useful to have some idea of the natural scales in relation to the units used in numerical solution of the equations. Firstly, we scale the reduced Planck mass, $8\pi G = \frac{1}{M_{pl}^2}$,  out of the Friedmann equation by rescaling the fields to be in Planck units: $\phi \rightarrow \phi/M_{pl}$, $\chi \rightarrow \chi/M_{pl}$, and absorbing factors of $M_{pl}^2$ into $B$, $D$ and the densities of the standard $\Lambda$CDM components. Next we change time variables to work in units of $H_0$: $\tau\rightarrow H_0 \tau$. This can be divided through the equations of motion and absorbed into the parameters and densities, so that the densities are now: $\rho \rightarrow \rho/(H_0^2 M_{pl}^2)$, and $\rho_i(a)/3 = \Omega_i(a)$. Finally we express all the parameters in the potential in Planck units, natural for a string inspired model, as $X \rightarrow M_{pl}^x X$, with $X$ the parameter and $x$ its mass dimension. Thus, finally we have:
\begin{align}
B &\rightarrow \left ( \frac{M_{pl}^2}{H_0^2} \right ) B \nonumber \\
D &\rightarrow \left ( \frac{M_{pl}^2}{H_0^2} \right ) D \nonumber \\
M^2 &\rightarrow \left( \frac{M_{pl}^2}{f_a^2} \right ) \left ( \frac{M_{pl}^2}{H_0^2} \right ) \bar{M}^2 \nonumber \\
\bar{M} &= \sqrt{\frac{M_{\text{SUSY}}}{M_{pl}}} \nonumber
\end{align}
All the parameters on the left hand side are now dimensionless, and it is these that will be used when quoting results.

Using $M_{pl}^2/H_0^2 \sim 10^{120}$, $f_a \sim 10^{16}\unit{GeV}$, and Planck scale SUSY to give an upper bound on $\bar{M} = 1$ gives an idea for the approximately natural scales of all the parameters\footnote{TeV scale SUSY is obviously more attractive from the point of view of resolving the hierarchy problem. However, $B$ and $D$ should also be set at the same magnitude, e.g. from potentials due to gaugino condensation, and our conclusions only rely on $(f_a^2/M_{pl}^2)M^2\sim B\sim D$.}:
\begin{align}
B &\lesssim 10^{120} \nonumber \\
D &\lesssim 10^{120} \nonumber \\
M &\lesssim 10^{62} \nonumber
\end{align}

For the fields, a large value of $\phi > 1$ will be transplanckian, and any value $\phi > \pi (f_a/M_{pl})$ will represent a departure from the periodic nature of the axion, whereas $\chi$ is an area/volume, so a large value in Planck units is not problematic, and is in fact what one expects in a LARGE volume scenario for moduli stabilisation. The initial conditions on the fields are free parameters in the model: $\chi$ representing a position in the landscape at the end of inflation, and $\phi$ being selected by spontaneous symmetry breaking. The flatness condition fixes the density in the axion-modulus system. Choosing the densities in $\Lambda$ and standard CDM to be close to their observed values allows the axion-modulus system to be set as sub-dominant. Thus for now we ignore the question of fixing the initial conditions so as to obtain $H_0$ at its observed value, since the subdominant components will not cause much variation away from this ($H_0 = 1$ in our units). The appropriate initial conditions for a universe containing a significant fraction in an axion-modulus component will be the subject of future work.

We expect the dimensionless parameters $C$ and $\tilde{C}$ to both be $\mathcal{O}(1)$.

\subsection{Comments}

At this stage some comments on the system of equations in relation to other models in the literature will be useful. A brief comparison will be made to three models: \cite{cervantes-cota2010,catena2007,beyer2010}, stating the main similarities and differences. The take-home message, though, will be that these models, whilst interesting, are connected to very different sources of new physics. They are for the most part motivated by scalar-tensor theories modifying the gravity sector to build dark energy models, whilst the model presented here is firmly cast in the context of HEP, strings, and the landscape, with dark matter candidates and modifications to the dark energy sector a by-product useful for phenomenology.

In \cite{cervantes-cota2010} the scalar field analogous to our modulus is coupled non-minimally to gravity via the term $\phi^2 R$, and has its own scalar potential. The field is also used to Higgs the Dirac dark matter particle via a Yukawa coupling $\bar{\psi}f(\phi)\psi$, thus introducing a $\phi$ dependent mass to the dark matter. However no explanation for this coupling is given in terms of fundamental particle physics (general couplings of this form can, however, be constructed in scalar-tensor theories by transforming between the Jordan and Einstein frames: see for example \cite{pettorino2008}). Assuming they have a standard neutralino dark matter particle, then the scalar field will be a Higgs of some supersymmetric extension to the Standard Model and one may worry about effects both on the freeze out mechanism for the neutralino, or on other effects where the true Higgs rolls and causes mass changes in the Standard Model particles. This model building issue aside, the authors go on to explore the effects of different potentials and coupling terms that allow matching to the standard cosmology. 

There are some important distinctions between the model of \cite{cervantes-cota2010} and the one presented here, the first being that the axion-modulus model makes no change in the gravity sector, and as such the FRW equations take their standard form, though \cite{cervantes-cota2010} can be brought into standard form via a conformal transformation: this is the usual degeneracy of Modified Gravity theories to scalar field theories of dark energy (again, see \cite{pettorino2008}). Secondly, the dark matter sector in \cite{cervantes-cota2010} is, apart from the coupling, assumed standard, whereas the axiverse model is motivated by non-standard dark matter components. Finally, many scalar potentials and couplings are explored in \cite{cervantes-cota2010} within a general framework, whereas in the axion-modulus model the forms are essentially fixed via the motivation in string theory, and would be formally fixed in any specific string realisation of the model. The motivation of \cite{cervantes-cota2010} is to explore late time dynamics of dark energy; the motivation here is to investigate dynamics in the matter sector and finds dark energy dynamics as a by product. Thus both models explore the same idea of dark sector couplings and non-standard dynamics, the model of \cite{cervantes-cota2010} being more phenomenologically general for cosmologists and dark energy phenomenologists, the model here being more theoretically  and phenomenologically general from a HEP point of view.

Coupling axions and moduli bears much similarity to ``axion-dilaton cosmology'' \cite{catena2007,sonner2006}. In these models the axion has no potential of its own but is coupled in its kinetic term to the dilaton after a conformal transformation renders the gravity sector standard: $\mathcal{L}\subset -\frac{1}{2}e^{\mu\sigma}(\partial \chi)^2$, where $\sigma$ is the dilaton, and $\chi$ the axion. The dilaton potential is then also exponential $\Lambda e^{-\lambda\sigma}$, appearing on the cosmological constant term after the conformal transformation. This model is then entirely constrained with only two free parameters. The dynamics in the scalar field sector consists of the existence of attractors bringing the equation of state periodically into accelerating phases, with possibly observable consequences as explored in \cite{linder2010}. The highly constrained axion-dilaton model should serve as a useful heuristic guide when thinking about the coupled scalar field dynamics of the axion-modulus model, the important differences being that axion-modulus model has an axion mass term with a coupling on it, but standard kinetic term, and the potential for the modulus has a finite field, negative potential minimum, in contrast to the dilaton. We will expect this to make quantitative and qualitative changes to the equation of state evolution.

Finally the model of \cite{beyer2010} bears the most resemblance to the axion-modulus model with the mapping of axion $\phi$ to ``geon'' $\chi$, and modulus $\chi$ to ``cosmon'' $\varphi$, the only important difference being that the modulus potential has a minimum, whilst the cosmon potential, being a dilaton, does not. The analysis of \cite{beyer2010} will accordingly be very useful for helping us understand the axion-modulus system. However, I choose to include a standard cosmological constant as well as the modulus, for the place holding reasons mentioned in Section~\ref{intro}, and consider this good practice. Whether or not $\Lambda$ can be removed from the axion-modulus system, with all the dark energy given from the scalar fields, will be the subject of future work, and if possible may have important consequences in relation to string theory. I also choose to include a standard dark matter component because we know that if ultralight scalar fields rolling on the time scales of interest here and in \cite{beyer2010} suppress structure formation according to their fraction \cite{marsh2010,amendola2005}, then, just like massive neutrinos, they cannot make up all of the dark matter, as is assumed in \cite{beyer2010}.

This concludes the discussion of the model.

\section{Results}
\label{results}

\subsection{Example Cosmology 1}
\label{example1}

\begin{figure}
\centering
\includegraphics[scale=0.4]{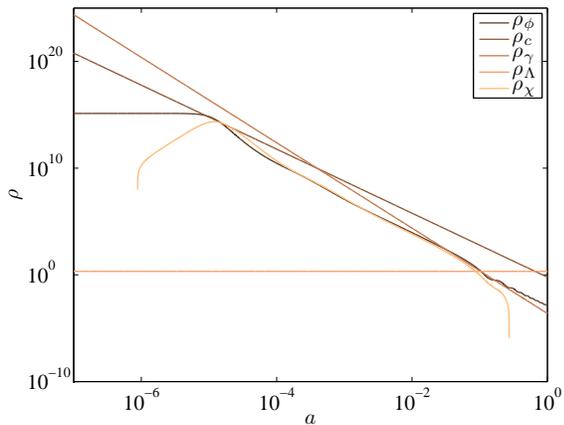}
\caption{Densities as a function of scale factor for the parameters of Equation~\ref{eqn:ex1params}. Notice the tracking dynamics of the axion-modulus system between $10^{-6} \lesssim a \lesssim 10^{-1}$, when the modulus gains positive energy density, and the end of this tracking at $a\gtrsim 10^{-1}$ caused when the axion field begins oscillations and the axion density makes its standard transition to CDM-like behaviour.}
\label{fig:ex1densities}
\end{figure}

\begin{figure}
\centering
\includegraphics[scale=0.4,trim=30mm 0mm 10mm 0mm,clip]{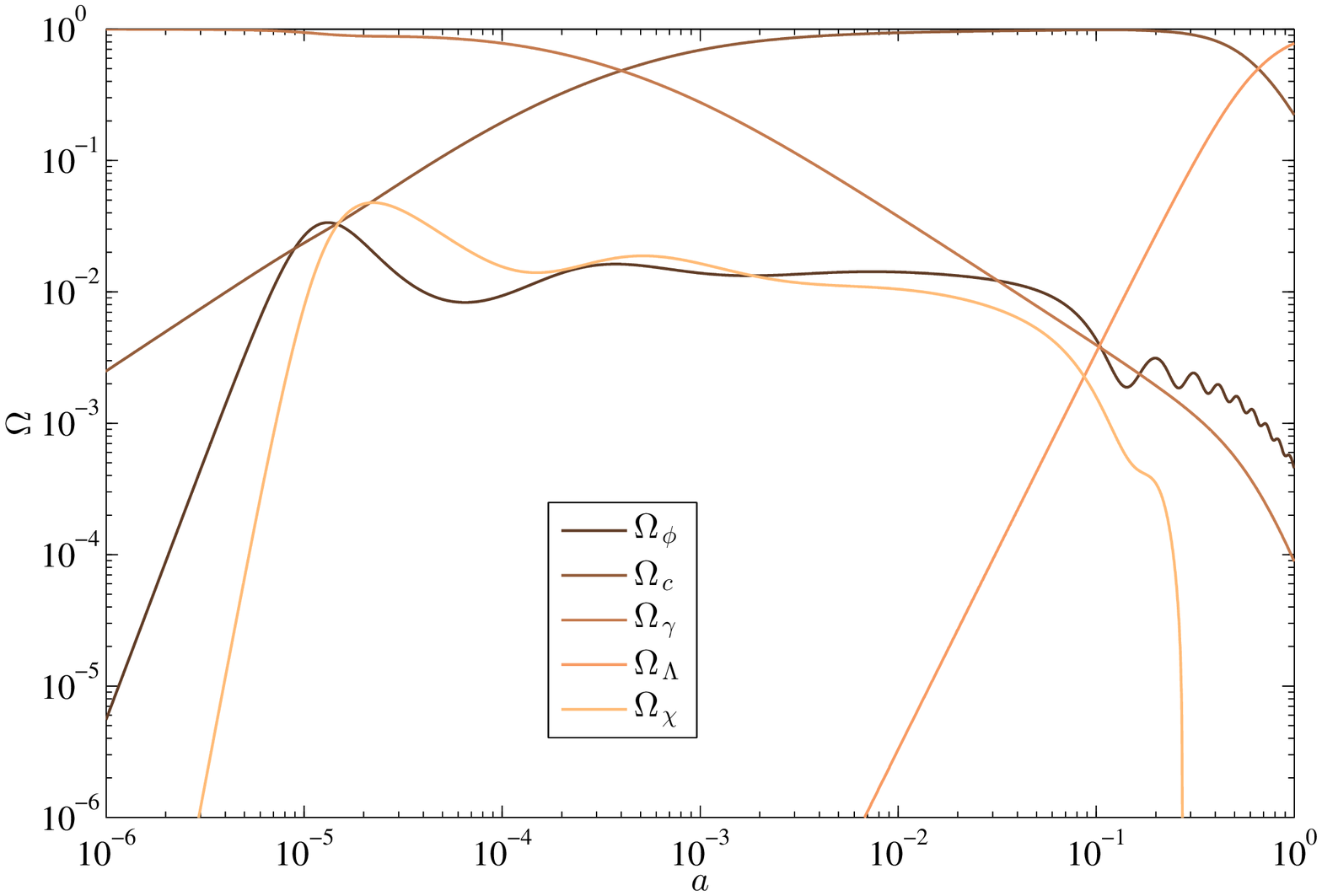}
\caption{Fraction of the critical density, $\Omega$, as a function of scale factor for the parameters of Equation~\ref{eqn:ex1params}.}
\label{fig:ex1omegas}
\end{figure}

This first example is concerned with parameter values close to those considered natural in Section~\ref{model}. The evolution of the various components of the density $\rho$ obtained from numerically integrating the equations of motion are plotted as a function of scale factor $a$ in Figure~\ref{fig:ex1densities}, with the associated $\Omega$'s plotted in Figure~\ref{fig:ex1omegas} for parameter values:
\begin{align}
M &= 10^{62} \nonumber \\
B &=10^{120} \nonumber \\
D &=10^{119} \nonumber \\
C &= \tilde{C} = 10 \nonumber \\
\phi_i &= 1\nonumber \\
\chi_i &= 25 \nonumber \\
\Omega_\Lambda &= 0.7 \nonumber \\
\Omega_c &= 0.2 \nonumber \\
\Omega_\gamma &= 8 \times 10^{-5} \nonumber \\
\label{eqn:ex1params}
\end{align}

\begin{figure}
\centering
\includegraphics[scale=0.4]{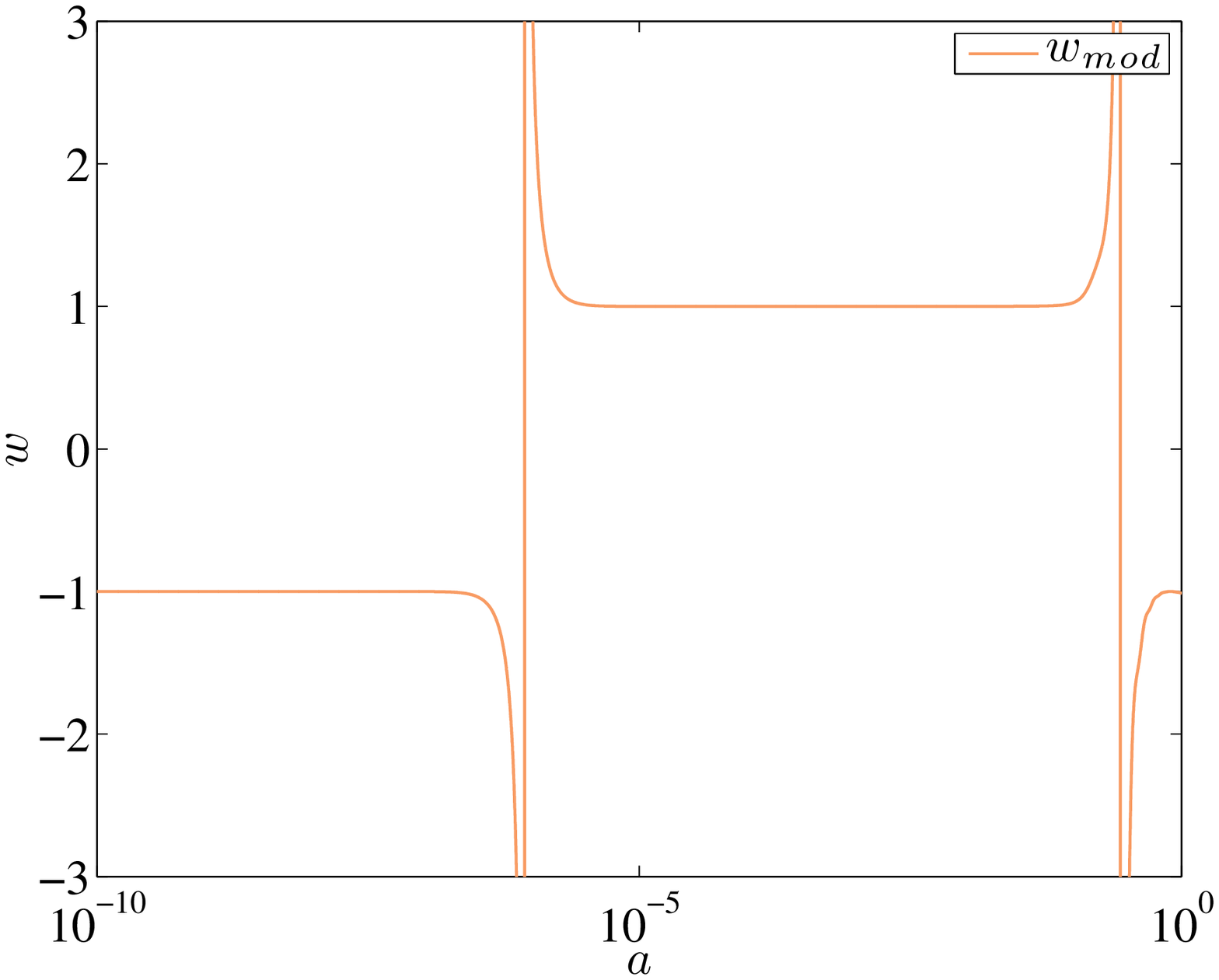}
\caption{Evolution of the modulus equation of state $w$ as a function of scale factor for the parameters of Equation~\ref{eqn:ex1params}.}
\label{fig:ex1wmod}
\end{figure}

\begin{figure}
\centering
\includegraphics[scale=0.4]{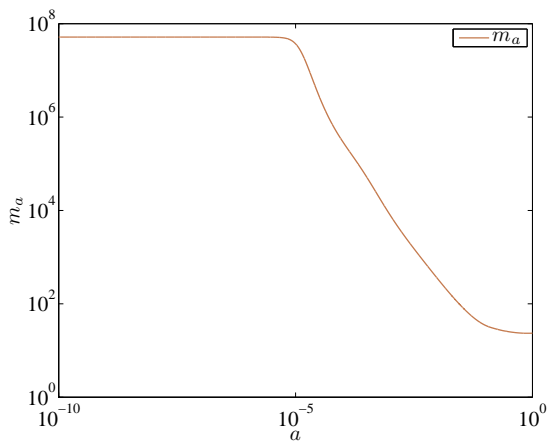}
\caption{Evolution of the axion mass as a function of scale factor for the parameters of Equation~\ref{eqn:ex1params}. Compare this to Figure~\ref{fig:ex1wmod}: the rolling occurs whilst $w_{mod}=1$ and the modulus has kinetic energy.}
\label{fig:ex1ma}
\end{figure}

In Figure~\ref{fig:ex1densities} there are a number of qualitative features worthy of comment. Firstly, the logarithmic scale prevents us from showing the small negative energy density associated to the modulus at early and late times, we only see it emerge onto the plot between scale factors $10^{-6}\lesssim a \lesssim 10^{-1}$ as it enters an attractor scaling solution. During this time the modulus is displaced from its initial value and its equation of state becomes kinetic dominated, $w = 1$, the additional energy density being kinetic, as demonstrated in Figure~\ref{fig:ex1wmod}. This transit causes a corresponding evolution of the axion mass, as demonstrated in Figure~\ref{fig:ex1ma}. It also induces a tracking behaviour in the axion density whilst the modulus is rolling. The axion field then begins its usual oscillatory behaviour and the axion behaves as dark matter for $a \gtrsim 10^{-1}$. 

\begin{figure}
\centering
\includegraphics[scale=0.4]{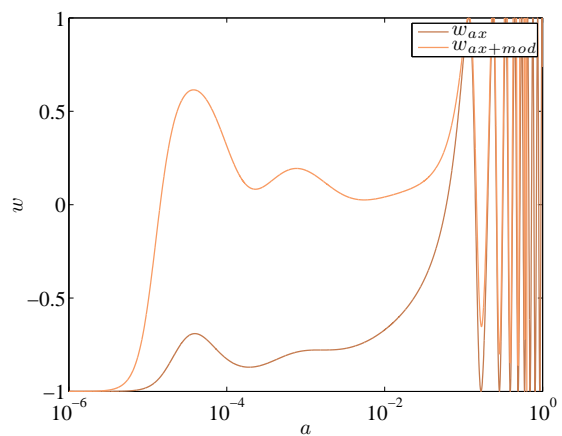}
\caption{Evolution of the axion and combined axion-modulus equation of state $w$ as a function of scale factor for the parameters of Equation~\ref{eqn:ex1params}. Notice the tracking behaviour of the combined equation of state, and the end of this when axion field oscillations cause oscillations in the pressure, averaging to zero.}
\label{fig:ex1ws}
\end{figure}

The effects of this are best viewed in terms of the equations of state for the combined systems. The axion equation of state, $w_{ax}$, and the axion-modulus equation of state, $w_{ax+mod}$, are shown in Figure~\ref{fig:ex1ws}. The axion equation of state differs from the usual case of a quick transition between $w = -1$, and oscillations averaging to $w = 0$. It is in the combined equation of state that we see tracking behaviour as $w$ tries to follow the equation of state of the dominant component, before the axion oscillations begin, which are the cosmic trigger event that destroys tracking and restabilises the modulus (in this context, stabilisation is defined by $w_{mod}\rightarrow -1$). Once the axion field begins oscillations it causes the equation of state to oscillate and the axion pressure averages to zero, i.e. the axion behaves as pressureless dark matter. The final cosmology today at $a = 1$ has a negligible negative component of energy density in the frozen modulus, whilst an ultralight dark matter axion makes up a fraction of the critical density of order that in radiation.

\begin{figure}
\centering
\includegraphics[scale=0.4]{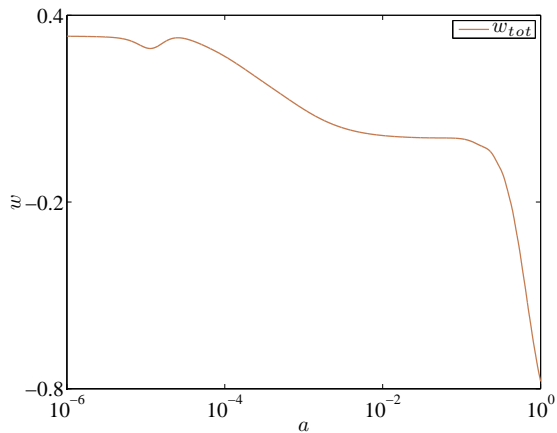}
\caption{Evolution of the total equation of state $w$ as a function of scale factor for the parameters of Equation~\ref{eqn:ex1params}. Note the appearance of a feature around $a \simeq 10^{-5}$ that departs from the standard smooth $\Lambda$CDM evolution. Compare this to Figure~\ref{fig:ex1omegas}, where an overshoot as the axion-modulus system enters its tracking solution causes a significant contribution to the critical density at this scale factor.}
\label{fig:ex1wtot}
\end{figure}

The resulting behaviour of the total equation of state for all fluids is plotted in Figure~\ref{fig:ex1wtot}, where we see that it develops a kink around scale factor $a \simeq 10^{-5}$ away from its usual $\Lambda$CDM evolution through matter-radiation equality, caused by the presence of a significant component of the fractional density due to the axion-modulus system at this time, as demonstrated in Figure~\ref{fig:ex1omegas}. This is EDE-like behaviour, although the equation of state is not dragged low enough to cause an early period of acceleration.

Having identified the main features of cosmology in this example, we now turn to briefly assess the dependency of these features on the parameters. A full description of the system in this way with its various degeneracies will require analysing it as a system of autonomous equations in the phase plane and the identification of the fixed points \cite{copeland1998,amendola2000}, which is left for a future work.

\begin{figure}
\centering
\includegraphics[scale=0.4]{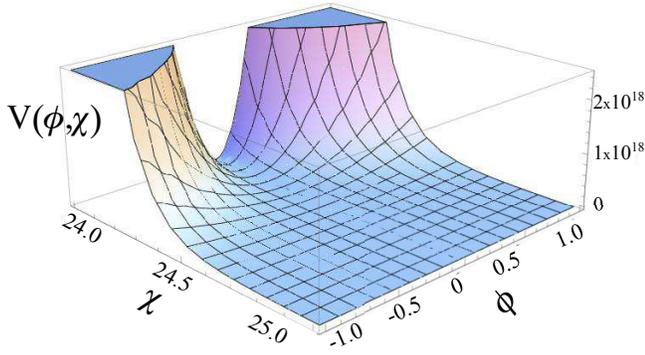}
\caption{The potential $V(\phi,\chi)$ of Equation~\ref{eqn:lagrangiantotal}, plotted for $\chi$ in the range $\{ 24,25\}$, $\phi$ in the range $\{ -1,1\}$, for the parameters of Equation~\ref{eqn:ex1params}.}
\label{fig:potential1}
\end{figure}

\begin{figure}
\centering
\includegraphics[scale=0.4]{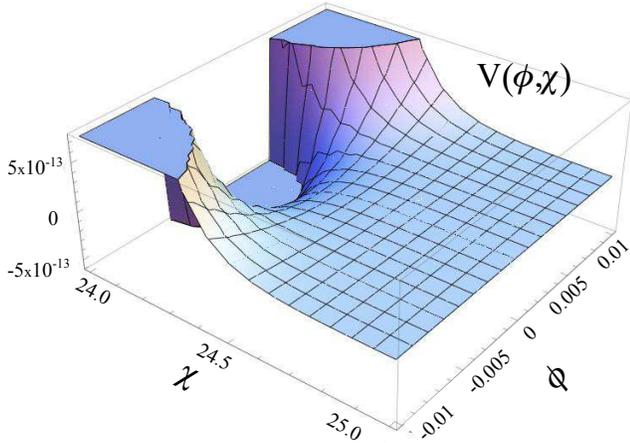}
\caption{The potential $V(\phi,\chi)$ of Equation~\ref{eqn:lagrangiantotal}, plotted for $\chi$ in the range $\{ 24,25\}$, $\phi$ in the range $\{ -0.01,0.01\}$, for the parameters of Equation~\ref{eqn:ex1params}.}
\label{fig:potential2}
\end{figure}

Decreasing the axion initial value, $\phi_i$, can cause a significant change in the modulus behaviour. With $C = \tilde{C}$ the condition for the modulus to have a finite real minimum is given by:
\begin{equation}
2D > M^2 \phi^2
\nonumber
\end{equation}
For the values used in Equation~\ref{eqn:ex1params} this can only occur as the axion decays, and the $\phi = 0$ minimum is at $\chi = 0.3$. Now, the dependence on axion initial condition can be most easily seen by plotting the potential $V(\phi,\chi)$. In Figure~\ref{fig:potential1} the potential is plotted for $\phi$ in the range $\{-1,1\}$, whilst in Figure~\ref{fig:potential2} it is plotted with $\phi$ in the range $\{-0.01,0.01 \}$. We see that, for small amplitude axion oscillations the exponential descent into the modulus minimum for $\phi = 0$ shows up strongly, whereas this feature is hidden on the same scale when the oscillations of the axion have a larger amplitude. We can similarly remove the axion from the spectrum almost entirely by increasing $\tilde{C}$ by an order of magnitude, making the axion effectively massless and allowing the modulus to rapidly reach its minimum.

Thus it is the axion oscillations which here stabilise the modulus away from its true minimum (as is visible comparing Figures~\ref{fig:ex1densities} and \ref{fig:ex1wmod}). If the modulus falls into this minimum then a large negative potential is generated causing the universe to collapse \cite{felder2002}. That this occurs prior to $a=1$ for a small axion initial field value corresponding to reasonable initial misalignment $\theta_i = 1$ (not shown) rules out these particular parameters, and will require tuning in these models to avoid it. This decay and collapse may occur in the future as the amplitude of axion oscillations decays \cite{marsh2010} for any model that looks viable today, and the parameters can then be used to estimate the lifetime of the universe \cite{wang2004}. The analysis of a collapsing universe in this model will be the subject of future work. Some further comments on entry into a collapsing phase will be made in Section~\ref{example2}.

Even a sight decrease in $\tilde{C}$ to $8$ or $9$ also has a dramatic effect on the potential, allowing for the appearance of new minima as the axion field oscillates about zero, and the $M^2$ and $D$ terms in the potential play off against one another. These new minima are sharp and highly localised in field space. Entry into them occurs as the modulus grows through its tracking solution and axion oscillations decay. A brief dip through them leads to a short period of negative potential domination, which the Friedmann equation is incapable of dealing with, and as such the analysis of this region of parameter space is left for a future work.


\subsection{Example Cosmology 2}
\label{example2}

\begin{figure}
\centering
\includegraphics[scale=0.4]{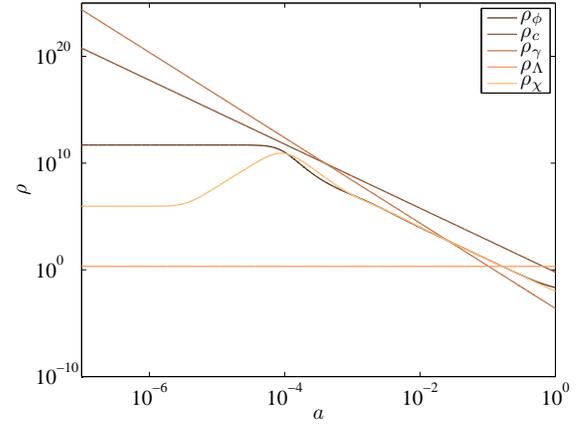}
\caption{Densities as a function of scale factor for the parameters of Equation~\ref{eqn:ex2params}.}
\label{fig:ex2densities}
\end{figure}

\begin{figure}
\centering
\includegraphics[scale=0.4,trim=18mm 0mm 10mm 0mm,clip]{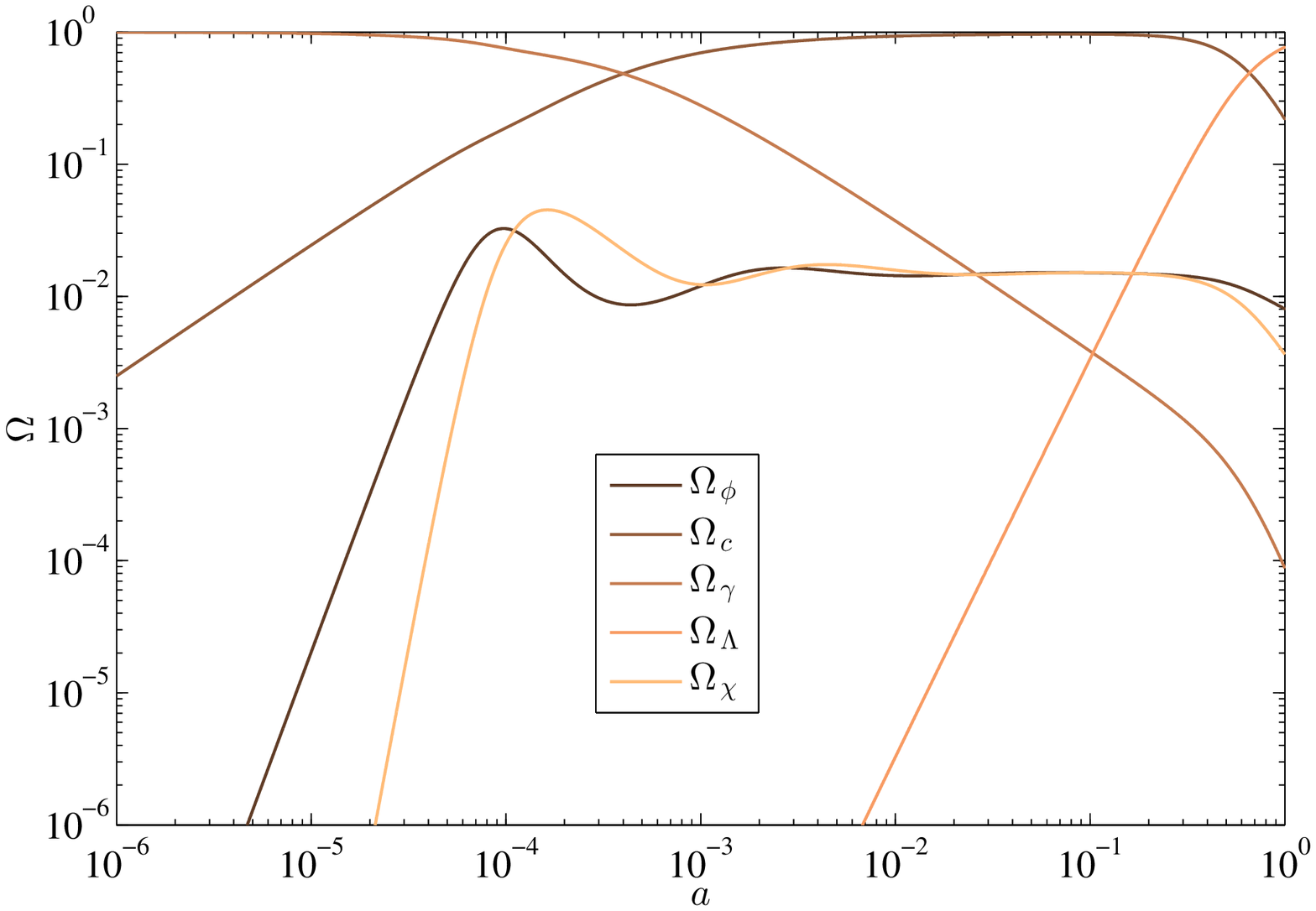}
\caption{Fraction of the critical density, $\Omega$, as a function of scale factor for the parameters of Equation~\ref{eqn:ex2params}.}
\label{fig:ex2omegas}
\end{figure}

\begin{figure}
\centering
\includegraphics[scale=0.4]{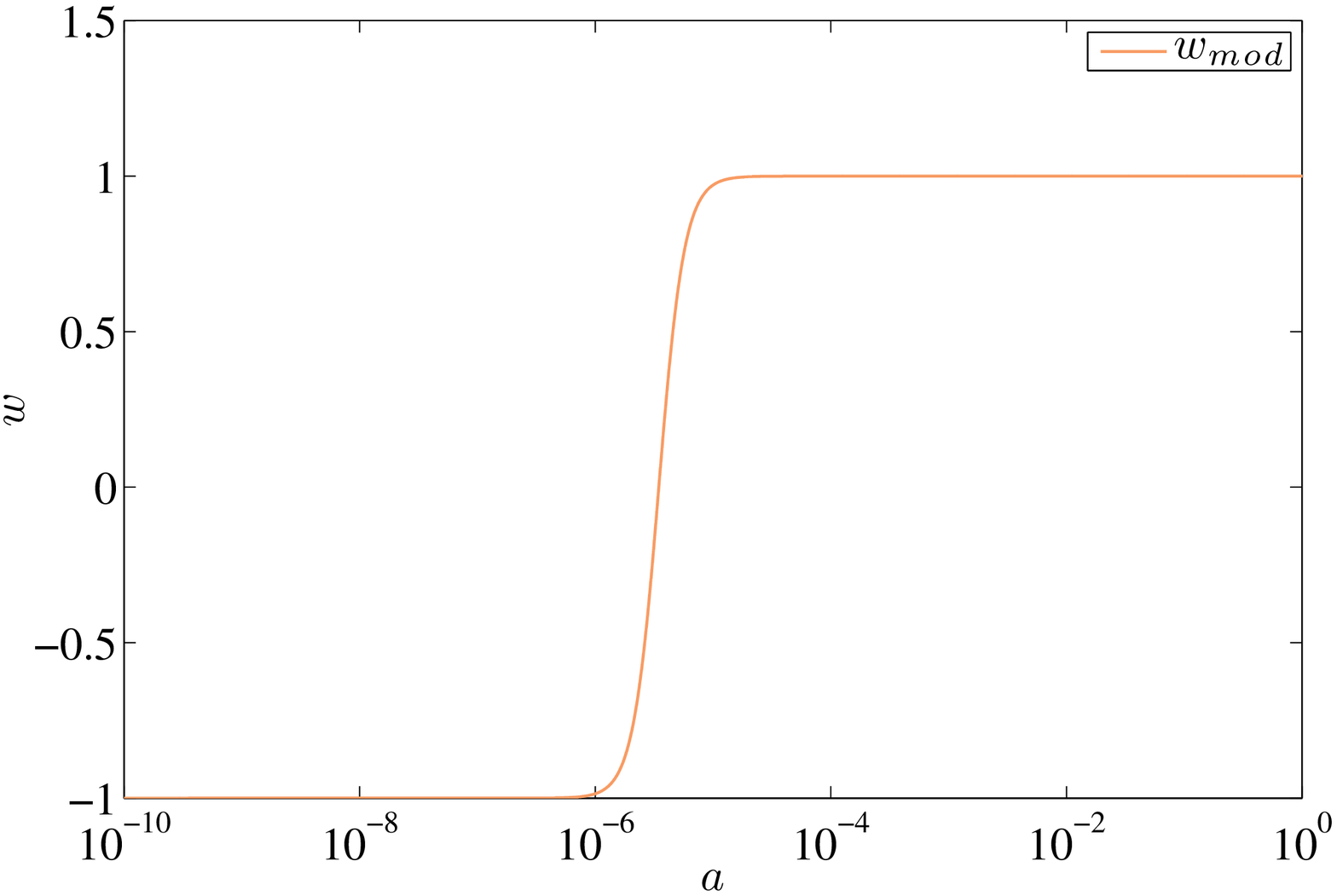}
\caption{Evolution of the modulus equation of state $w$ as a function of scale factor for the parameters of Equation~\ref{eqn:ex2params}.}
\label{fig:ex2wmod}
\end{figure}

\begin{figure}
\centering
\includegraphics[scale=0.4]{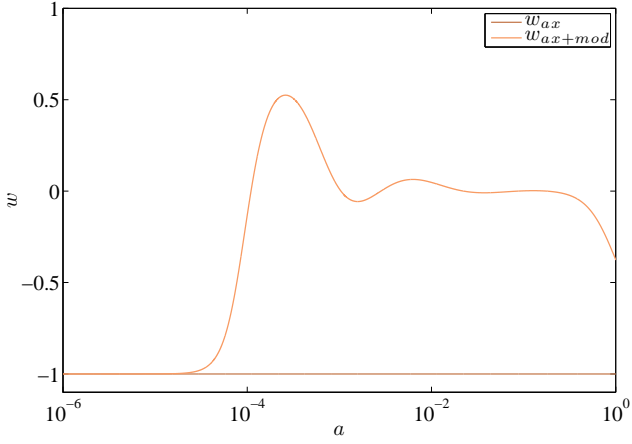}
\caption{Evolution of the axion and combined axion-modulus equation of state $w$ as a function of scale factor for the parameters of Equation~\ref{eqn:ex2params}.}
\label{fig:ex2ws}
\end{figure}

\begin{figure}
\centering
\includegraphics[scale=0.4]{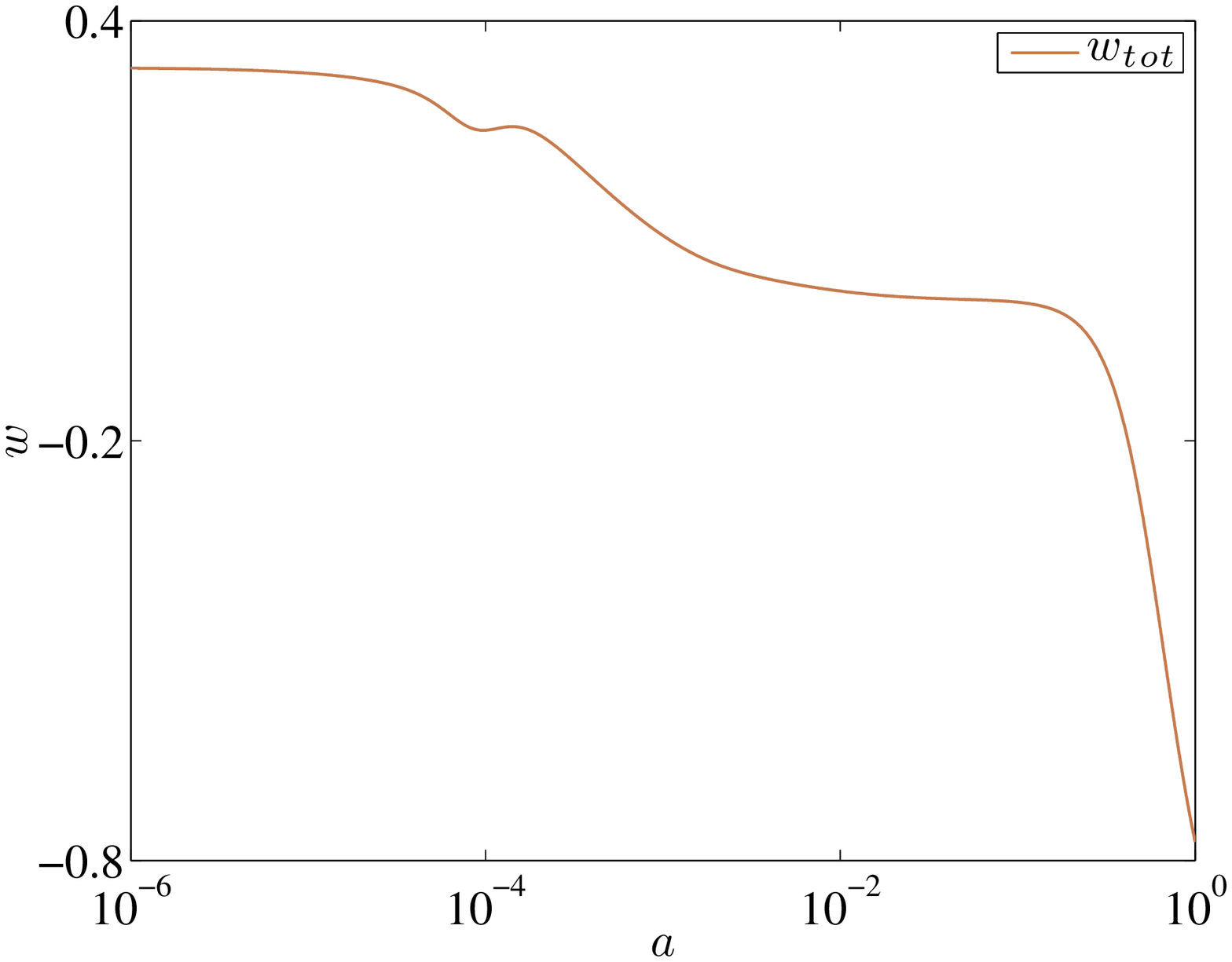}
\caption{Evolution of the total equation of state $w$ as a function of scale factor for the parameters of Equation~\ref{eqn:ex2params}.}
\label{fig:ex2wtot}
\end{figure}

\begin{figure}
\centering
\includegraphics[scale=0.4]{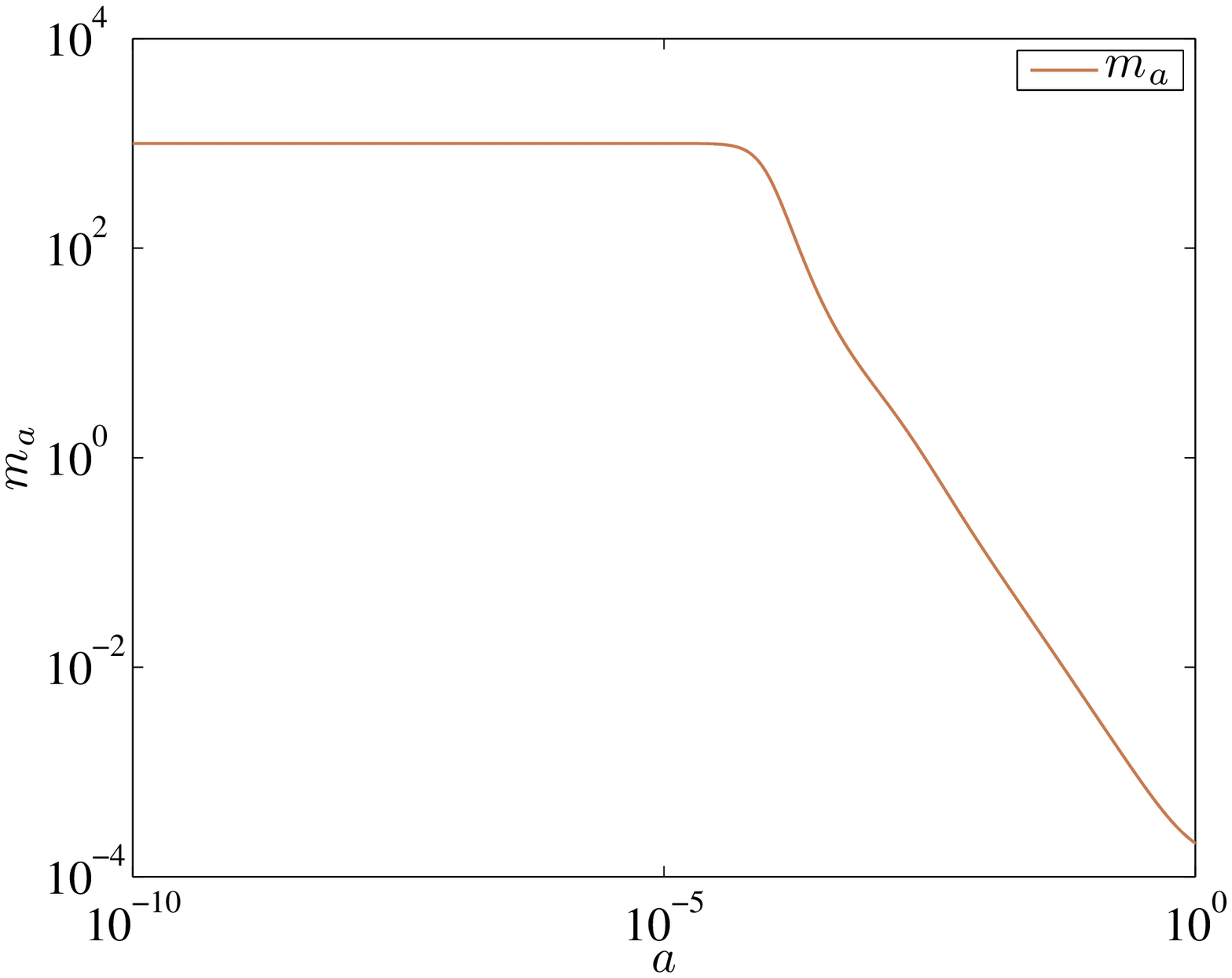}
\caption{Evolution of the axion mass as a function of scale factor for the parameters of Equation~\ref{eqn:ex2params}.}
\label{fig:ex2ma}
\end{figure}

\begin{figure}
\centering
\includegraphics[scale=0.4]{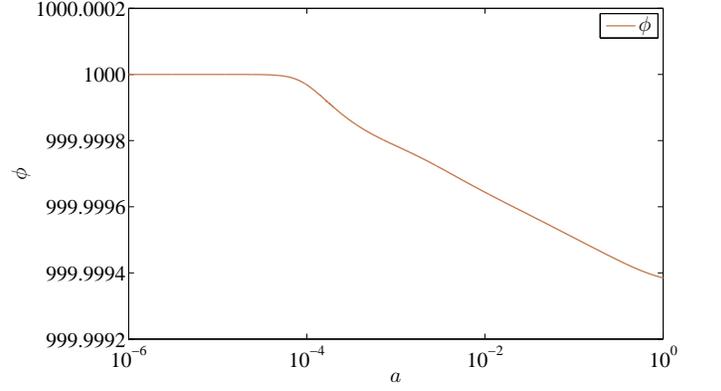}
\caption{Evolution of the axion field as a function of scale factor for the parameters of Equation~\ref{eqn:ex2params}.}
\label{fig:ex2phi}
\end{figure}

The purpose of this example is to demonstrate the large freedom in choosing values for parameters in these models by producing a similar and viable cosmology to that presented in Section~\ref{example1}, but with parameters many orders of magnitude different. This also shows that we expect many degeneracies in the parameters, with only the ratios of some being relevant. Specifically, the parameters in this example are:
\begin{align}
M &= 10^{3} \nonumber \\
B &=10^{6} \nonumber \\
D &=10^{5} \nonumber \\
C &= \tilde{C} = 10 \nonumber \\
\phi_i &= 10^3\nonumber \\
\chi_i &= 10^{-5} \nonumber \\
\Omega_\Lambda &= 0.7 \nonumber \\
\Omega_c &= 0.2 \nonumber \\
\Omega_\gamma &= 8 \times 10^{-5} \nonumber \\
\label{eqn:ex2params}
\end{align}
The evolution of the densities, $\Omega$'s, equations of state, axion mass, and the axion field, are shown in Figures \ref{fig:ex2densities}, \ref{fig:ex2omegas}, \ref{fig:ex2wmod}, \ref{fig:ex2ws}, \ref{fig:ex2wtot}, \ref{fig:ex2ma} and \ref{fig:ex2phi}. 


This example shows best the possibility of the axion-modulus system to display tracking EDE behaviour. The densities of both fields are always positive and show a scaling with the dominant component of energy density. Whilst the modulus equation of state is kinetic dominated once rolling begins, $w = 1$, the axion equation of state remains always potential dominated, $w = -1$, and the axion slowly rolls, never beginning oscillations as the mass is exponentially damped below the Hubble scale (see Figures~\ref{fig:ex2ma}, \ref{fig:ex2phi}). Tracking for the combined system thus persists into the present epoch, and should persist indefinitely, being absent the trigger event of axion oscillations to end it. In this case, the splitting of the energy density between the two components as done in Equation~\ref{eqn:rhosplit} is not really so clear and it makes sense to speak more in terms if the combined axion-modulus system as a quintessence fluid.

In this example the equation of state for the axion-modulus system (see Figure~\ref{fig:ex2ws}) has a novel shape, with no oscillations, varying through behaviour like a cosmological constant, radiation, matter, and quintessence as it begins rolling and goes through its various scaling stages. The total equation of state (Figure~\ref{fig:ex2wtot}) is again marginally perturbed by the presence of a significant axion-modulus component near equality, and the final fractional density in the axion-modulus system is $\Omega_{ax+mod} \simeq 0.01$. Compare to the previous example: the fractional density in the axion-modulus system is approximately the same (being fixed by the scaling solution by $C$), but makes a significant contribution at the slightly later time of $a \sim 10^{-4}$, and correspondingly alters the total equation of state at this time. This demonstrates that there is control in the parameters over features in the total equation of state. The density at late times in this example is larger, since the tracking is never ended by axion oscialltions.

Reducing the initial field for the axion in this example only changes the $\Omega$'s slightly, as one would expect when tracking is present, but moves the scale of the kink in $w_{tot}$ to yet later times (not shown). Increasing $M$ undoes this change, as we expect since for slowly varying $\phi$ it is the $\chi$ dependence of the potential that counts, and $M^2\phi^2$ is just a multiplicative factor in the exponential $\chi$ potential, like $B$ and $D$. It is the $M^2\phi^2$ term that is dominant for the parameters in question and as such the example is similar to the well studied case of a single scalar field with an exponential potential. 

\begin{figure}
\centering
\includegraphics[scale=0.4]{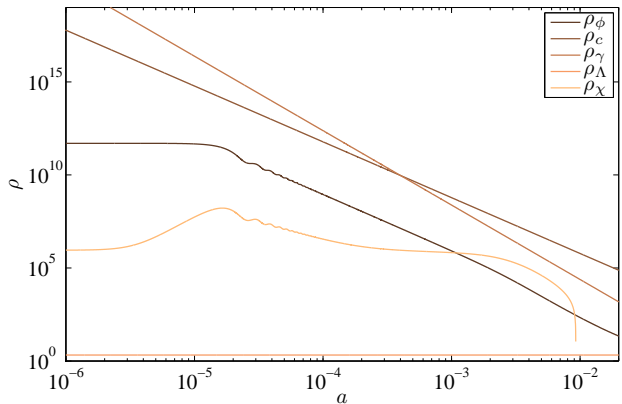}
\caption{Densities as a function of scale factor for the parameters of Equation~\ref{eqn:ex2params} but with the change $M \rightarrow 10^8$, $\phi_i \rightarrow 10^{-2}$. Notice that the axion field behaves completely as standard \cite{amendola2005,marsh2010}. The modulus begins at positive $\Lambda$ like behaviour, before a fall into a negative potential dominated phase, signalling the onset of cosmological collapse.}
\label{fig:ex2densitiesalt}
\end{figure}

\begin{figure}
\centering
\includegraphics[scale=0.4]{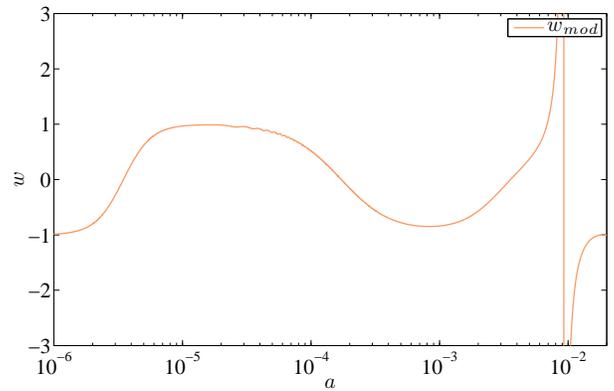}
\caption{Evolution of the modulus equation of state $w$ as a function of scale factor for the parameters of Equation~\ref{eqn:ex2params}, but with the change $M \rightarrow 10^8$, $\phi_i \rightarrow 10^{-2}$. The tracking solution is never quite found, as the equation of state makes a slow oscillation. When the modulus settles into its true negative potential minimum we see a the spike in the equation of state at $a\simeq 10^{-2}$}
\label{fig:ex2wmodalt}
\end{figure}

Further variations now cause qualitative changes in the cosmology, which are demonstrated in Figures~\ref{fig:ex2densitiesalt} and \ref{fig:ex2wmodalt}. A higher mass $M = 10^8$, and lower misalignment angle $\theta_i = 1$ ($\phi_i = 10^{-2}$) destroys the axion tracker and the axion behaves as standard, oscillating and making up a subleading fraction of dark matter, with the modulus potential dominated by the axion mass term and thus contributing a positive energy density. However, at later times the modulus falls into and bounces from its own potential minimum, eventually dominating with negative potential and again signalling rapid entry into a phase of cosmic contraction \cite{felder2002,wang2004}. During this evolution the modulus equation of state makes a slow oscillation as the field moves around in its potential and finds the true minimum.


We finally note that contraction cannot be properly analysed in the framework presented here, since the Friedmann equation alone does not allow for it. The framework to use is the Friedmann acceleration equation for $\ddot{a}$ \cite{wang2004}, but applying that to this system, although simple in principle, is left for a future work.

\section{conclusions}
\label{conclusions}

In this paper a model has been proposed that introduces a coupling between axions and moduli in the string axiverse. The coupling is motivated by the observation that the mechanism causing axion masses to distribute on a logarithmic scale, and thus for some to exist with masses in the range $10^{-33}\unit{eV}\lesssim m_a \lesssim10^{-18}\unit{eV}$ where their cosmological dynamics can produce interesting phenomenology, is due to an exponential dependence of the mass on the size of cycles in the compact space. The sizes of these cycles are controlled by scalar fields called moduli, which are themselves dynamical. Moduli stabilisation is an important problem in string theory. We investigate the possibility that if there are light axions that roll in their potentials on timescales of cosmological interest then some moduli may also roll in their potentials, given by a general form for stabilisation, and that this vacuum destabilisation can be affected in turn by the presence of the axions.

We then explore the consequences of having one cosmologically relevant axion and allowing its counterpart modulus to also roll. The resulting system, in terms of the background expansion of the universe, is a simple one of two coupled scalar fields with a scalar potential containing a number of free parameters, decoupled from the other cosmic fluids. The potential causes the fields to have scaling solutions where the energy density tracks that of the dominant component. This destabilises the modulus, and the resulting evolution causes an evolution of the axion mass, altering the dynamics from the most simple case of a single decoupled axion.

The axion-modulus system has a number of different phases in its evolution, of which we have identified some which may be of phenomenological interest. A common feature is that when tracking begins, the fraction of the total energy density in the axion-modulus system rises to an appreciable level and causes a non-standard evolution in the total equation of state in a manner similar to models of Early Dark Energy. In the examples considered in this paper this resulted in a decrease of the equation of state by $\mathcal{O}(10\%)$ around the epoch of equality.

The fate of the axion-modulus system, and consequently the fate of cosmic expansion, then depends on the rate of decay of the axion mass and the amplitude of axion oscillations. If the axion oscillates and the amplitude of its oscillations decrease below some critical value then the modulus falls into its globally stabilising minimum at negative potential, which if this potential comes to dominate the energy density will signal the onset of an epoch of cosmic collapse. If the dynamical axion mass can be sufficiently damped by the modulus evolution then oscillations cannot begin and the tracking solution remains stable.

There is also the possibility that a modulus that looks to have been stabilised by Hubble friction at early times might be destabilised when its counterpart axion begins to roll at late times, and that this vacuum destabilisation may be observable indirectly through its effects on couplings of both gravity and the standard model, though the viability of this scenario in a realistic model requires much further attention.

We have shown that these phenomena might be expected as fairly generic since they are exhibited for values of the parameters spanning many orders of magnitude. We also note that in a scenario such as the string axiverse, where there is a plethora of light fields at the phenomenologists disposal then it is possible to create cosmologies where phenomena like those described here may occur at multiple different epochs in the history and future of the universe, both by having multiple fields or by the two field dynamics spiralling towards some attractor and periodically entering and exiting different regions in phase space.

In the interests of simplicity no attempt was made to describe a universe where a significant and controllable fraction of the dark sector energy density is contributed by the axion-modulus system, except in the case where rapid decay of the modulus leads to cosmological collapse. This question of initial conditions will be the subject of a future work.

Other future work will focus on delineating the regions of parameter space which give rise to the phenomena described in this paper and thus allow one to both construct desirable models for cosmological phenomenology based in string theory, and also to place limits on the possible values of parameters in any model of this type, which will be severely limited by the possibility of cosmological collapse.

The coupling induces a tracking of energy densities in the axion-modulus system, which, just as it does for quintessence, may be fruitfully used to address problems of fine tuning, which are of a major concern for high $f_a$ axions. The tracking dynamics may also allow a modulus which appears naively to be stabilised in some negative energy density anti-de-Sitter minimum to in fact play the role of a cosmological constant leading to late time de-Sitter expansion, which may contribute in some way to the solution of the cosmological constant problem in string theory.

This simple extension of the axiverse has a rich structure and suggests models which we hope may be of use to both cosmologists and string theorists, and displays features which may, when fully investigated, be measured and constrained by cosmological experiments, particularly once the perturbations have been analysed to allow computation of effects in the cosmic microwave background and matter power spectrum.



\section*{Acknowledgements}
\vspace{-10pt}
I would like to thank Pedro Ferreira for much support throughout the project. I also thank Joe Conlon, Ed Copeland, Chris Gordon, Ren\'{e}e Hlozek, Francisco Pedro, John March-Russell and Ewan Tarrant for useful discussions throughout the course of this work. I acknowledge support from an STFC Postgraduate Studentship.


\bibliographystyle{apsrev}
\bibliography{doddyoxford}

\end{document}